\documentclass[aps,preprint]{revtex4-1}
\usepackage{graphicx}
\usepackage{amsmath}
\usepackage{makeidx}
\usepackage{amsfonts}
\usepackage{amssymb}

\begin{document}

\title{Optimization by a quantum reinforcement algorithm}

\author{A. Ramezanpour}
\email{aramezanpour@gmail.com}
\affiliation{Department of Physics, University of Neyshabur, Neyshabur, Iran}
\affiliation{Leiden Academic Centre for Drug Research, Faculty of Mathematics and Natural Sciences, Leiden University, Leiden, The Netherlands}
\date{\today}

\date{\today}

\begin{abstract}
A reinforcement algorithm solves a classical optimization problem by introducing a feedback to the system which slowly changes the energy landscape and converges the algorithm to an optimal solution in the configuration space. Here, we use this strategy to concentrate (localize) the wave function of a quantum particle, which explores the configuration space of the problem, preferentially on an optimal configuration. We examine the method by solving numerically the equations governing the evolution of the system, which are similar to the nonlinear Schrödinger equations, for small problem sizes. In particular, we observe that reinforcement increases the minimal energy gap of the system in a quantum annealing algorithm. Our numerical simulations and the latter observation show that such kind of quantum feedback might be helpful in solving a computationally hard optimization problem by a quantum reinforcement algorithm.  
\end{abstract}


\maketitle

\section{Introduction}\label{S0}
Reinforcement is a very useful technique in machine learning and optimization algorithms for the study of computationally hard optimization problems \cite{SB-book-1998,MM-book-2009}. The main idea is based on rewarding good decisions or modifying the energy landscape in a way that leads the algorithm to an optimal solution. This is usually done by introducing an appropriate feedback to the system, which depends on the information provided by the algorithm, to guide the optimization process. In this paper, we study a quantum reinforcement algorithm, which employs a continuous-time quantum random walk to explore the configuration space of an optimization problem. We show that such kind of quantum feedbacks can converge the quantum particle towards a solution by a preferential localization of the wave function in the solution space.        

Consider the problem of finding a solution to a classical optimization problem, identified by a probability distribution over the configuration space of the problem variables. We assume that the probability distribution is nonzero only for (a possibly large number of) configurations in the subspace of solutions. Then a decimation algorithm to find a solution works by fixing the value of a randomly chosen variable according to the marginal probability of that variable. The algorithm continues until the value of every variable is fixed. The marginal probabilities at each step are obtained by an approximate sampling algorithm, e.g., Monte Carlo, conditioned on the values of the already decimated variables. Instead of fixing the variables one by one, a reinforcement optimization algorithm modifies smoothly the joint probability distribution of the variables by changing slowly the values of some external local fields acting on the variables \cite{BZ-prl-2006}. These local fields use the estimated marginal probabilities of the variables to concentrate the joint probability distribution more and more on a single configuration in the solution space. 

On the other hand, a quantum algorithm exploits the computational power of a quantum system to solve a computationally difficult problem \cite{NC-book-2002,KS-book-2002,Shor-1994,Grover-1997}. Quantum random walks \cite{ALZ-pra-1993,EG-pra-1998,SKW-pra-2003,K-cp-2003,A-jqi-2003,S-ieee-2004} and adiabatic quantum computation \cite{F-sci-2001,KL-qc-2004} are important examples of quantum approaches to universal computations \cite{C-prl-2009,A-siam-2008}. Specifically, we should mention recent efforts in constructing effective shortcuts to adiabaticity \cite{B-jpa-2009,R-prl-2009,C-prl-2013}, nonunitary evolution of quantum random walks and non-Hermitian quantum annealing \cite{M-jcp-2008,NG-pra-2012}, and investigations of quantum annealing with nonstoquastic Hamiltonians \cite{I-prl-2016,H-arx-2016,SJN-pra-2017}. Another related study is the quantum reinforcement learning algorithm \cite{Rlearning-ieee-2008}, which is a quantum implementation of the reinforcement learning algorithm.     

The wave function of a quantum particle in the complex energy landscape of an optimization problem can undergo a localization transition, which may limit the efficiency of a quantum annealing algorithm \cite{AHJ-pnas-2010,BLPS-prb-2016}. Here, we propose a quantum reinforcement algorithm, which works by concentrating the wave function preferentially on the subspace of optimal configurations. The algorithm exploits the information provided by the instantaneous wave function of the system, or expectation values of some local observables, to steer the evolution of the quantum system. In addition, we show that such a quantum reinforcement can increase the minimum energy gap that the system encounters in a quantum annealing algorithm.

It is known that a nonlinear quantum mechanic can be exploited by a quantum computer to solve a computationally hard problem in a polynomial time \cite{lloyd-prl-1998}; this does not mean that quantum mechanics is nonlinear in nature, or any nonlinearity in the time evolution of the quantum system is computationally beneficial. Here, we show that a kind of nonlinear quantum evolution inspired from the classical reinforcement algorithms can be used to increase the energy gap and speedup the computation compared with the conventional quantum annealing algorithm.

\section{Main Definitions}\label{S1}
We consider the classical optimization problem of minimizing an energy function $E(\boldsymbol\sigma)$ of $N$ binary spins $\sigma_i=\pm 1$. To be specific, as the benchmark we take a (fully-connected) random spin model, with $E(\boldsymbol\sigma)=-\sum_{i<j}J_{ij}\sigma_i\sigma_j$.
The couplings $J_{ij}$ are independent Gaussian random variables of mean zero and variance $1/N$. The scaling is chosen to have an extensive energy of order $N$. This model is known as the Sherrignton-Kirkpatrick (SK) model \cite{SK-prl-1975},

We shall use a continuous-time quantum random walk to explore the space of spin configurations $\boldsymbol\sigma=\{\sigma_1,\cdots,\sigma_N\}$. The space is a hypercube of $2^N$ sites corresponding to the total number of spin configurations. The Hamiltonian of the particle in the energy landscape of the classical optimization problem is given by
\begin{align}
H=\sum_{\boldsymbol\sigma} E(\boldsymbol\sigma) |\boldsymbol\sigma\rangle\langle \boldsymbol\sigma|-\sum_{\boldsymbol\sigma}\sum_{i=1}^N\Gamma\left( |\boldsymbol\sigma^{-i}\rangle\langle \boldsymbol\sigma| + |\boldsymbol\sigma\rangle\langle \boldsymbol\sigma^{-i}| \right). 
\end{align}
The parameter $\Gamma$ determines the strength of tunneling from $|\boldsymbol\sigma \rangle$ to a neighboring state $|\boldsymbol\sigma^{-i} \rangle$. Here $|\boldsymbol\sigma^{-i} \rangle$ denotes the spin state, which is different from $|\boldsymbol\sigma \rangle$ only at site $i$. In terms of the quantum spin variables (Pauli matrices), the above Hamiltonian reads as $H=-\sum_{i<j}J_{ij}\sigma_i^z\sigma_j^z-\sum_i \Gamma \sigma_i^x$. Additionally, the basis states $|\boldsymbol\sigma\rangle$ are the $N$-spin states with definite $\sigma_i^z$ values, that is $\sigma_i^z|\boldsymbol\sigma\rangle=\sigma_i|\boldsymbol\sigma\rangle $.  
Starting from an initial state $|\psi(0)\rangle$, time evolution of the system is governed by the Schroedinger equation, $\hat{i}\frac{d}{dt} |\psi(t)\rangle = H |\psi(t)\rangle$ with $\hbar=1$.

\section{Quantum Reinforcement Algorithm}\label{S2}
The goal here is to find a solution to the classical optimization problem by following the time evolution of a quantum system. A quantum annealing (QA) algorithm \cite{F-sci-2001} starts from the ground state of $H_x\equiv -\sum_i \Gamma \sigma_i^x$ and changes slowly the Hamiltonian to $H_c \equiv -\sum_{i<j}J_{ij}\sigma_i^z\sigma_j^z$. The adiabatic theorem then ensures that in the absence of level crossing, the system follows the instantaneous ground state of the time dependent Hamiltonian $H_{QA}(t)= s(t)H_c+ [1-s(t)]H_x$.  
The annealing parameter $s(t)$ changes slowly from zero at $t=0$, to one at $t=T$. For instance, in a linear annealing schedule $s(t)=t/T$.    

In a quantum reinforcement (QR) algorithm, we add a reinforcement term to the Hamiltonian which favors the spin states of higher probability.
More precisely, the Hamiltonian is $H_{QR}(t)= H_c+ H_x+H_r(t)$, where the reinforcement term reads as follows:  
\begin{align}
H_r(t)\equiv -r(t)\sum_{\boldsymbol\sigma}|\psi(\boldsymbol\sigma;t)|^2 |\boldsymbol\sigma\rangle\langle \boldsymbol\sigma|.
\end{align}
The reinforcement parameter $r(t)$ is zero at the beginning and grows slowly in magnitude with time. In a linear reinforcement schedule we take $r(t)=(t/T)2^Nr_0$. In other words, as the time passes, the on-site energy at state $|\boldsymbol\sigma \rangle$ decreases with an amount that is proportional to the probability of finding the walker at that site $|\psi(\boldsymbol\sigma;t)|^2$. This probability could be exponentially small at the beginning of the process. That is why here we scale the reinforcement parameter with $2^N$. To have an extensive Hamiltonian for $N\to \infty$, one can also change the scaling with time such that $r(t) \propto N$ at the end of the process, where the wave function is of order one.   

For comparison with the QA algorithm, we also study a reinforced quantum annealing (rQA), where 
\begin{align}
H_{rQA}(t) \equiv s(t)[H_c+H_r(t)]+ [1-s(t)]H_x.
\end{align}
This allows us to examine the effect of reinforcement on the behavior of the quantum annealing algorithm.
The reinforcement parameter here is a constant $r(t)=Nr_0$. So, the time dependence of $H_r(t)$ is determined by $\psi(\boldsymbol\sigma;t)$, which can safely be replaced by the instantaneous ground state of the system for an adiabatic process.  
 
To obtain a local version of the above algorithms, we first replace the $|\psi(\boldsymbol\sigma;t)|^2$ with $\ln |\psi(\boldsymbol\sigma;t)|^2$, which is an increasing function of the probability distribution. Note that we can always write $|\psi(\boldsymbol\sigma;t)|^2=\exp(\sum_i K_i\sigma_i+\sum_{i<j}K_{ij}\sigma_i\sigma_j+\cdots)/Z$, taking into account all the possible multi-spin interactions; $Z$ is the normalization constant. The coupling parameters $K_i, K_{ij}, \dots$ in principle can be determined from the expectation values $\langle \sigma_i^z\rangle, \langle \sigma_i^z\sigma_j^z\rangle, \dots$. A local quantum reinforcement (lQR) algorithm then is obtained by approximating the wave function with a product state,  
\begin{align}
H_r^{local}(t) \equiv -r(t) \sum_{\boldsymbol\sigma}\sum_i K_i\sigma_i|\boldsymbol\sigma\rangle\langle \boldsymbol\sigma|.
\end{align}
The reinforcement fields $K_i$ depend on the average spin values $m_i=\sum_{\boldsymbol\sigma} \sigma_i |\psi(\boldsymbol\sigma;t)|^2$ through $K_i=\frac{1}{2}\ln((1+m_i)/(1-m_i))$. Here we increase the reinforcement parameter with time as $r(t)=r_0t$.
More accurate approximations of the wave function and the local quantum reinforcement algorithm can be obtained by considering the two-spin and the higher-order interactions in the expansion.    

\begin{figure}
\includegraphics[width=16cm]{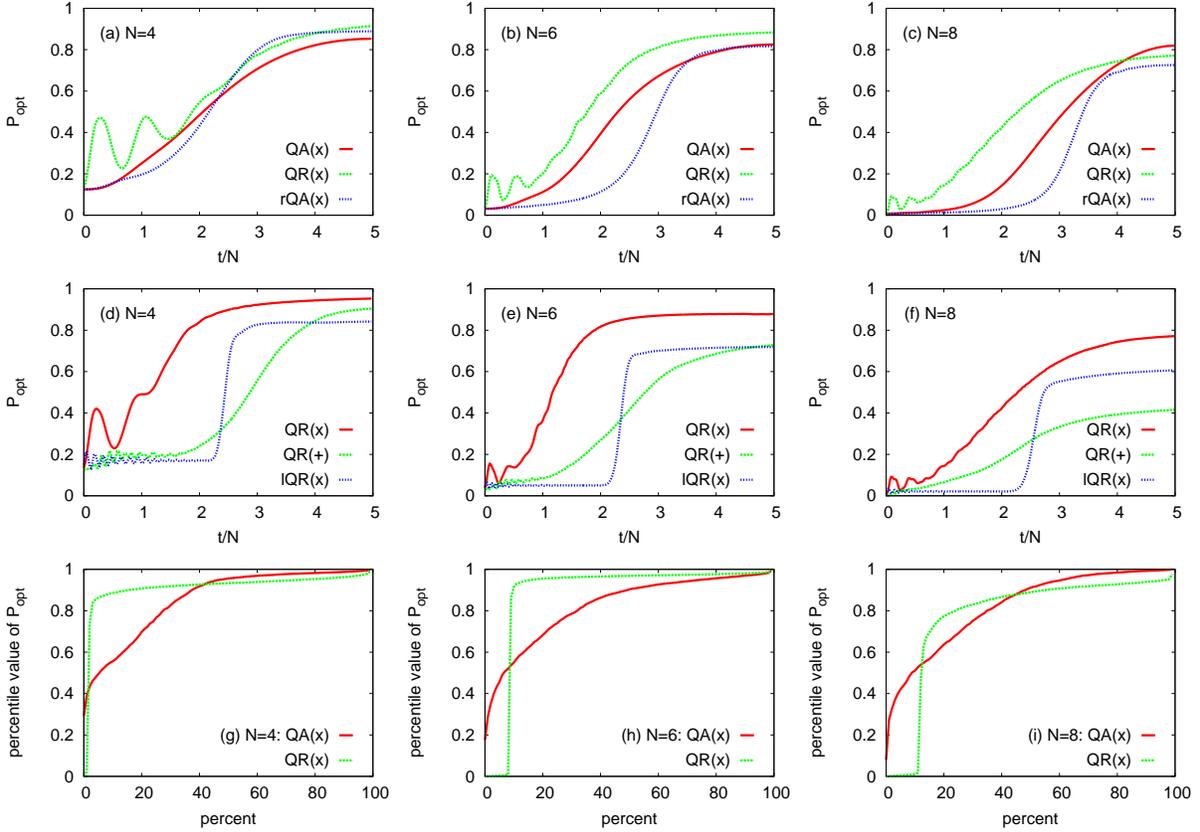} 
\caption{Success probability $P_{opt}$ of the algorithms for different number of spins $(N)$ in the SK model. Panels (a)-(b)-(c) show the results obtained by the quantum annealing (QA), quantum reinforcement (QR), and reinforced quantum annealing (rQA) algorithms. Panels (d)-(e)-(f) compare the results obtained by the QR algorithm for different initial conditions and the local quantum reinforcement (lQR) algorithm. Panels (g)-(h)-(i) display the percentile values of the success probability at the end of the process ($t=T$). The initial states are indicated by (x) for the equal superposition of all spin configurations, and (+) for the all-positive spin configuration. The data are results of $2000$ independent realizations of the random spin model. The statistical errors are about $0.01$. For the QA(x) we used $\Gamma=0.3$ (a,b,g,h) and $\Gamma=0.5$ (c,i). For the QR(x) we used $\Gamma=0.8, r_0=1$ (a,g), $\Gamma=1.2, r_0=1$ (b,h), $\Gamma=1.3, r_0=0.5$ (c,f,i), $\Gamma=1, r_0=2.5$ (d), and $\Gamma=1.5, r_0=2.5$ (e). For the rQA(x) we used $\Gamma=0.5, r_0=1$ (a), $\Gamma=0.9, r_0=1$ (b), and $\Gamma=1.2, r_0=1$ (c). For the QR(+) we used $\Gamma=3, r_0=3.5$ (d), $\Gamma=2.5, r_0=2.5$ (e), and $\Gamma=2.5, r_0=2.0$ (f). For the lQR(x) we used $\Gamma=3, r_0=1$ (d), $\Gamma=3.5, r_0=0.5$ (e), and $\Gamma=2.5, r_0=0.2$ (f).} \label{f1}
\end{figure}

\section{Results and Discussion}\label{S3}
Figure \ref{f1} shows the success probability $\sum_{\boldsymbol\sigma_{opt}}|\langle \boldsymbol\sigma_{opt} | \psi(t) \rangle|^2$ of the above algorithms for the SK model. The initial state $|\psi(0)\rangle$ is the equal superposition of spin states $|x\rangle=\frac{1}{\sqrt{2^N}}\sum_{\boldsymbol\sigma}|\boldsymbol\sigma\rangle$, or the all-positive state $|\boldsymbol{+}\rangle=|++\cdots +\rangle$. In each case we tried different values of the parameters to obtain roughly the best performances. As expected for a quantum random walk, the algorithms are sensitive to the initial state of the system \cite{K-cp-2003}. We see from the figure that the QR algorithms can localize a considerable fraction of the wave function on the optimal spin configurations. Nevertheless, the performance of these algorithms degrades by increasing the number of spins (for $3<N<9$). We know that the success probability of the QA algorithm decreases exponentially with $N$ because of the exponentially small energy gaps of the Hamiltonian \cite{AHJ-pnas-2010}. On the other hand, we know from Ref. \cite{lloyd-prl-1998} that nonlinear quantum evolution could be helpful; therefore, it would be interesting to see how the success probability of the QR algorithm scales with the number of spins for larger systems. 

\begin{figure}
\includegraphics[width=8cm]{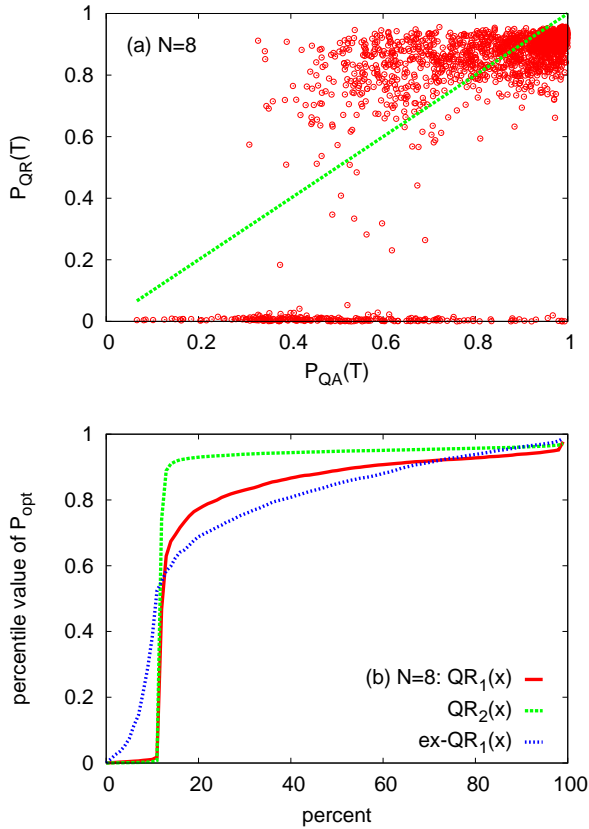} 
\caption{Success probabilities $P_{QR}, P_{QA}$ of the QR and QA algorithms in the SK model. (a) $P_{QR}(T)$ vs $P_{QA}(T)$ at the end of the process $t=T$ for $2000$ independent realizations of the problem. The parameters and initial conditions are similar to the ones in Fig. \ref{f1}(c). (b) Comparing the percentile values of the success probability in the one- and two-stage quantum reinforcement algorithms (QR$_1$, QR$_2$) starting from the equal superposition of all the spin states. The parameters in the two stages are $r_1(t)= 0.5(t/T)2^N, \Gamma_1=1.3$ and $r_2(t)=r_1(t)+0.15, \Gamma_2=\Gamma_1$ with $T=5N$ for each stage. Here ex-QR$_1$ denotes the one-stage QR algorithm with parameters $r(t)=0.5(t/T)(1-t/T)2^N, \Gamma(t)=2(1-t/T)$, which go to zero at the end of the process. The data are results of $2000$ independent realizations of the random spin model.} \label{f2}
\end{figure}

Figure \ref{f1} also shows the percentile values of the success probability for some independent realizations of the problem; we see that there are always a finite fraction of the problem instances for which the success probability of the QR algorithm is nearly zero; we can indeed enhance the success probability for these instances by slightly changing the algorithm parameters. In Fig. \ref{f2}(a) we compare the success probability of the QR algorithm with that of the QA algorithm for a number of independent problem instances. We observe that the QR algorithm displays large sample to sample fluctuations, with very good or very bad performances compared to the QA algorithm. 

The good point with the QR algorithms is that we do not need the ground state of the Hamiltonian to initialize these algorithms. Therefore, one can restart the algorithm at any stage of the evolution with an arbitrary wave function, or a spin configuration, which is sampled from the wave function of the system at that moment (see also Refs. \cite{guzik-qip-2011,mqa-njp-2017}). Figure \ref{f2}(b) gives the success probability of a two-stage QR algorithm where the final wave function of the first stage is taken for the initial state in the second stage of the algorithm.  In this study we used simple reinforcement schedules [$r(t)\propto t$ and $\Gamma(t)=\mathrm{const}$], which is not necessarily the best way of exploiting the reinforcement; in general, the reinforcement parameter is expected to grow at the beginning and the tunneling may diminish at the end of the process. 
Figure \ref{f2}(b) shows also the success probability of the one-stage QR algorithm for the case in which the reinforcement parameter increases with time at the beginning of the process and then goes to zero along with the hopping parameter $\Gamma$. This is to demonstrate that we do not need an exponentially large reinforcement when the wave function is of order one.

\begin{figure}
\includegraphics[width=8cm]{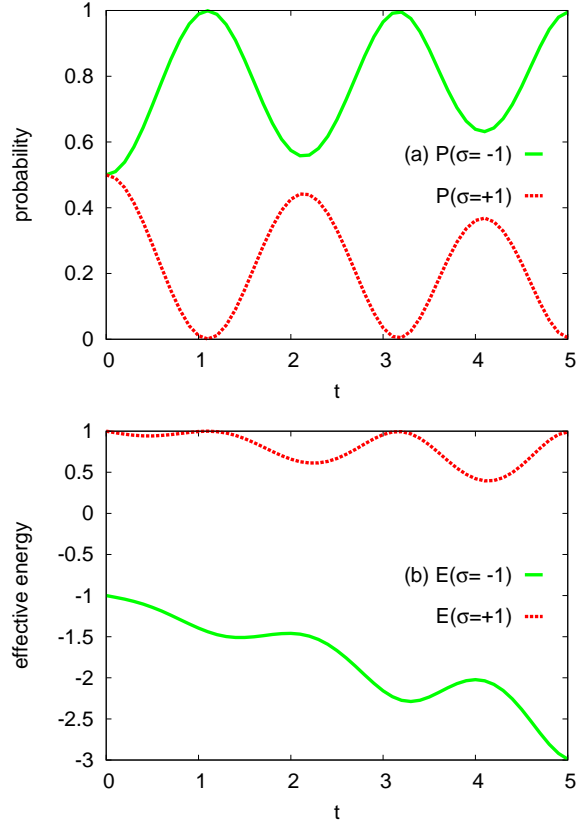} 
\caption{Time evolution of the two-level system in the quantum reinforcement algorithm. (a) The probabilities $|\psi(-;t)|^2$ and $|\psi(+;t)|^2$, and (b) the effective energy $E(\sigma;t)=h\sigma-r(t)|\psi(\sigma;t)|^2$. The parameters are $h=\Gamma=1$, and $r(t)=2(t/T)$ with $T=5$.}\label{f3}
\end{figure}

\begin{figure}
\includegraphics[width=16cm]{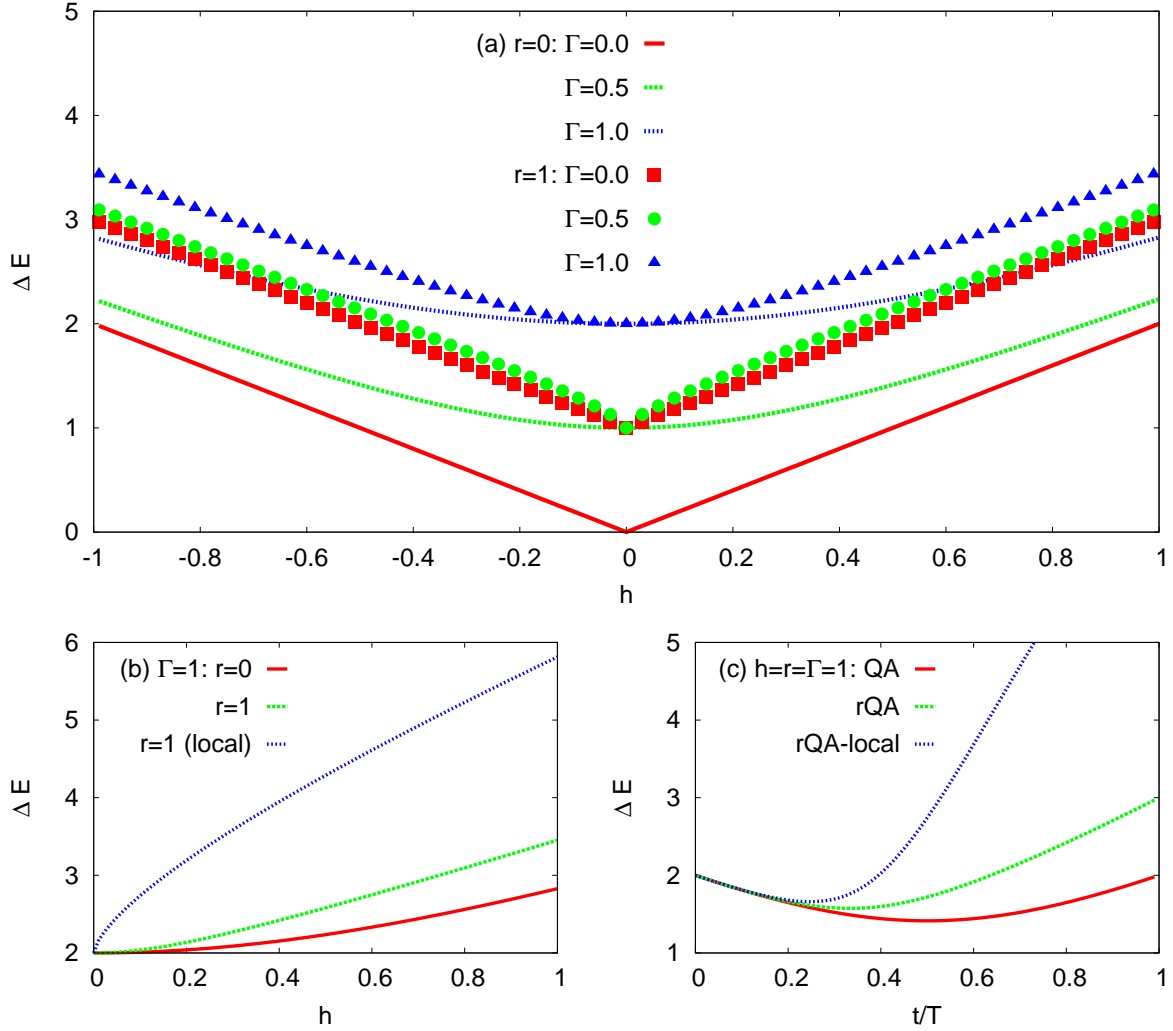} 
\caption{(a) The energy gap of the two-level system $\Delta E(h)$ for different values of $\Gamma$ and $r$ in the reinforced Hamiltonian $H_R^{(2)}$. (b) Energy gap of the local reinforced Hamiltonian for given values of $\Gamma$ and $r$. (c) Time dependence of the energy gap $\Delta E(t)$ for given values of $h, \Gamma$, and $r$ in the quantum annealing (QA), reinforced quantum annealing (rQA), and the local version of reinforced quantum annealing (rQA-local).}\label{f4}
\end{figure}

To see what happens close to a level crossing, we consider a two-level system with energy function $E(\sigma)=h\sigma$, where $\sigma=\pm 1$. For the initial state we take $|\psi(0)\rangle=(|-\rangle+|+\rangle)/\sqrt{2}$. In the QR algorithm, the evolution is governed by the following Hamiltonian,
\begin{align}
H_{QR}^{(2)}(t)=\left(\begin{array}{cc} 
h-r(t)|\psi(+;t)|^2 & -\Gamma \\
-\Gamma & -h-r(t)|\psi(-;t)|^2
\end{array}\right).
\end{align}
Figure \ref{f3} shows the time dependence of the wave function and the effective (reinforced) energies $E(\sigma;t)\equiv h\sigma-r(t)|\psi(\sigma;t)|^2$. We see that the energy landscape is favoring more and more the ground state of the system as the time passes. Notice also the oscillations in the wave function and the effective energies; these are reminiscent of the oscillations that are observed in the amplitude amplification algorithm in search problems \cite{B-cm-2002}.       

Next, we consider the two-level system with a reinforced Hamiltonian 
\begin{align}
H_{R}^{(2)}=\left(\begin{array}{cc} 
h-r|\psi_0(+)|^2 & -\Gamma \\
-\Gamma & -h-r|\psi_0(-)|^2
\end{array}\right),
\end{align}
where $\psi_0(\sigma)$ is the ground state. In Fig. \ref{f4}(a) we see the energy gap of the system for different values of the parameters. The eigenvalues and eigenstates of the Hamiltonian have been obtained numerically by an iterative algorithm.  
In Fig. \ref{f4}(b), we compare the energy gap of the above reinforced Hamiltonian with that of a local reinforced Hamiltonian, where $|\psi_0(\sigma)|^2$ is replaced with $K\sigma$. As before, the coupling is given by $K=\frac{1}{2}\ln[(1+m)/(1-m)]$, where $m=|\psi_0(+)|^2-|\psi_0(-)|^2$ is the magnetization. In the reinforced quantum annealing (rQA), the Hamiltonian of the two-level system reads as follows:
\begin{align}
H_{rQA}^{(2)}(t)=\left(\begin{array}{cc} 
(t/T)[h-r|\psi_0(+;t)|^2] & -(1-t/T)\Gamma \\
-(1-t/T)\Gamma & (t/T)[-h-r|\psi_0(-;t)|^2]
\end{array}\right).
\end{align}
Figure \ref{f4}(c) displays the time dependence of the energy gap for the quantum annealing and the (local) reinforced quantum annealing algorithms. As the figure shows, in both the cases the reinforcement tends to increase the energy gap of the two-level system.

\begin{figure}
\includegraphics[width=16cm]{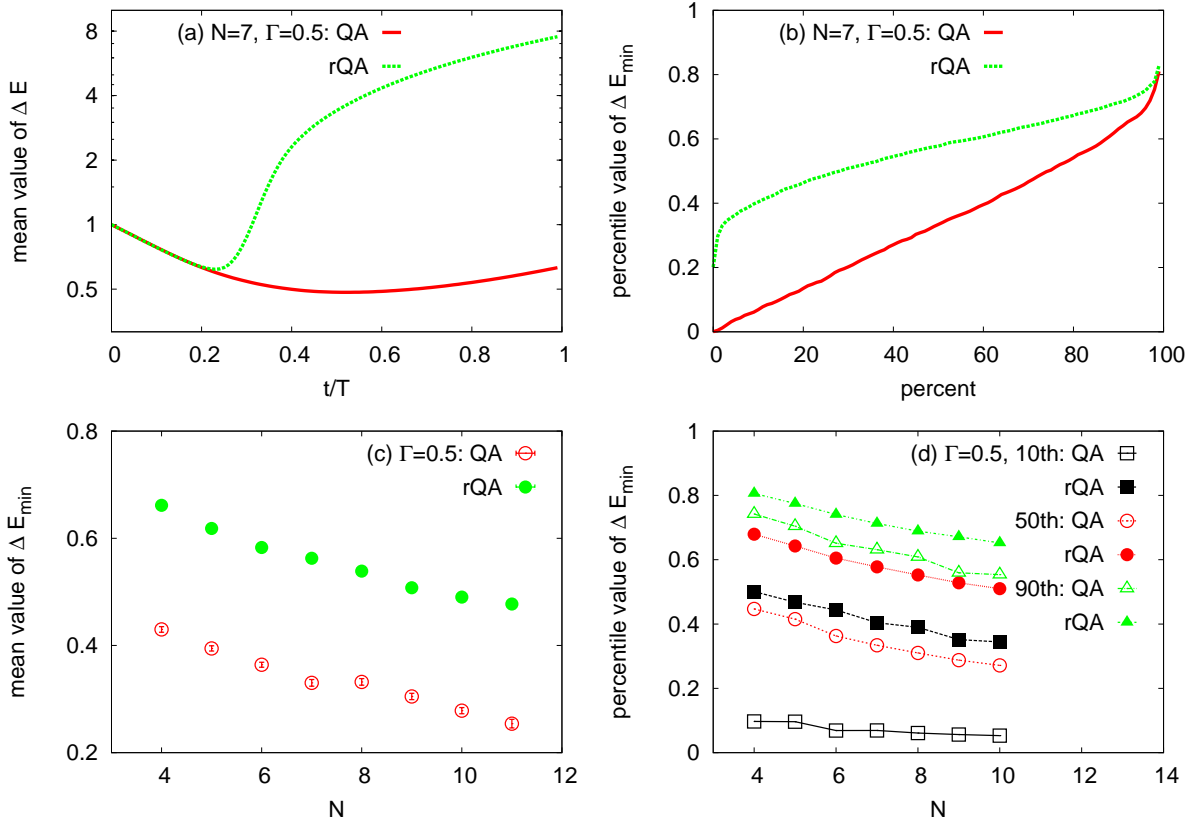} 
\caption{The energy gap of the SK model in presence of random fields in the quantum annealing (QA) and reinforced quantum annealing (rQA) with $r_0=1$. (a) The average energy gap $\Delta E(t)$ vs the evolution time, and (b) the percentile values of the minimum energy gap $\Delta E_{min}$, for a given number of spins $N$ and $\Gamma$. (c) The average of the minimal energy gap encountered in the annealing process, and (d) the percentile values of $\Delta E_{min}$, vs the number of spins. The data are results of $2000$ independent realizations of the random model.}\label{f5}
\end{figure}

In Fig. \ref{f5} we check the above observation for larger random spin systems. Here, for the classical problem we take the SK model with random fields, $E(\boldsymbol\sigma)=-\sum_{i<j}J_{ij}\sigma_i\sigma_j-\sum_i h_i\sigma_i$. The additional interactions with the external fields ensure that the ground state is not degenerate. The fields $h_i$ are independent Gaussian random variables of mean zero and variance one. In the numerical simulations we assume the system follows the instantaneous ground state of the Hamiltonian $H_{rQA}(t)$. As the figure shows, the reinforcement increases the energy gap and can be useful as another heuristic algorithm to improve the efficiency of the conventional quantum annealing algorithm.

\section{Conclusion}\label{S4}
In summary, our numerical simulations of random spin systems show that quantum reinforcement algorithms might be useful in solving a computationally expensive optimization problem. Clearly, more studies are required to see how this strategy works in larger problem sizes. In particular, the local version of the quantum reinforcement algorithm (introduced in Sec. \ref{S2}) can be studied by a quantum Monte Carlo algorithm for simulation of an open quantum system at sufficiently small temperatures, to examine the method for larger systems. Another challenge lies in the experimental realization of such quantum feedbacks in practice. Recent advances in quantum control theory \cite{DK-pra-1999,Qestimation-prl-2006}, e.g., the concept of continuous measurement of a quantum system, could be helpful in this direction. A naive approach is to approximate the nonlinear evolution of the quantum system by a sequence of estimations of the quantum state followed by linear evolutions of the quantum state \cite{Qcontrol-book}. The reinforcement terms in the Hamiltonian are updated only in the estimation stage, the time evolution of the system is then controlled by the updated Hamiltonian in the evolution stage.
Specifically, in the case of the local QR algorithm, we need to engineer a local Hamiltonian with couplings that can be estimated from a weak measurement of $N$ commuting spin variables, perhaps via interaction with some ancillary spins.



\end{document}